\documentstyle[twoside,fleqn,npb,epsf]{article}
\def\Pom{{\bf I\!P}}

\def\lsim{\mathrel{\rlap{\lower4pt\hbox{\hskip1pt$\sim$}}
    \raise1pt\hbox{$<$}}}         

\def\gsim{\mathrel{\rlap{\lower4pt\hbox{\hskip1pt$\sim$}}
    \raise1pt\hbox{$>$}}}         

\newcommand{\AmS}{{\protect\the\textfont2
  A\kern-.1667em\lower.5ex\hbox{M}\kern-.125emS}}

\title{Spin dependence of diffractive DIS }

\author{N.N. Nikolaev\address{Institut f. Kernphysik, Forschungszentrum J\"ulich,
D-52450 J\"ulich, Germany\\
 \& L.D.Landau Institute for Theoretical Physics,
142432 Chernogolovka, Russia}%
        \thanks{This work has been partly supported by the grant 
INTAS-96-0579}}

\begin{document}

\begin{abstract}
I review the recent progress in the theory of $s$-channel helicity 
nonconservation (SCHNC) effects
in diffractive DIS including the unitarity driven demise of 
 the Burkhardt-Cottingham sum rule and strong scaling departure from 
the Wandzura-Wilczek relation.
\end{abstract}

\maketitle


The motto of high energy QCD is the quark helicity conservation which is
often believed to entail the $s$-channel helicity conservation (SCHC)  
at small $x$. Here I review the recent work 
(\cite{NPZLT,KNZ98,IN99,AINP,G2},  which shows this belief is groundless and there
is an extremely rich helicity-flip physics at small $x$.

The backbone of DIS  is the Compton scattering (CS) $\gamma^{*}_{\mu}p\to 
\gamma^{*'}_{\nu}p'$,
which at small-$x$ can be viewed as a (i) dissociation $\gamma^{*}\to q\bar{q}$ 
followed by (ii) elastic scattering $q\bar{q} p \to q\bar{q} p'$ with {\bf exact}
conservation of quark helicity and (iii) fusion $q\bar{q} \to gamma^{*'}$. 
The CS amplitude $A_{\nu\mu}$ can be written as 
\begin{equation}
A_{\nu\mu}=\Psi^{*}_{\nu,\lambda\bar{\lambda}}\otimes A_{q\bar{q}}\otimes
\Psi_{\mu,\lambda\bar{\lambda}}
\label{eq:1}
\end{equation}
where $\lambda,\bar{\lambda}$ stands for $q,\bar{q}$ helicities,
$\Psi_{\mu,\lambda\bar{\lambda}}$ is the wave function of the $q\bar{q}$ 
Fock state of the photon. The $q\bar{q}$-proton scattering
kernel $A_{q\bar{q}}$  does not depend on, and conserves, the $q,\bar{q}$
helicities and is calculable in terms
of the gluon structure function of the target proton $G(x,\overline{Q}^{2})$ taken
at $\bar{x}={1\over 2}(Q^{2}+m_{V}^{2})/(Q^{2}+W^{2})$ and a certain 
hard scale $\overline{Q}^{2}$, see below. 

For
nonrelativistic massive quarks, $m_{f}^{2} \gg Q^{2}$, one only has
transitions with $\lambda +\bar{\lambda}=\mu$. However, 
the relativistic P-waves give rise to 
transitions
of transverse photons into the $q\bar{q}$ state with $\lambda +\bar{\lambda}=0$
in which the helicity of the photon as transferred to the $q\bar{q}$ orbital 
angular momentum. This state $\lambda +\bar{\lambda}=0$ is shared by the 
transverse (T) and longitudinal (L) photons
which entails \cite{NPZLT} 
the  $s$-channel helicity nonconserving (SCHNC) single-helicity 
flip $T\to L$ which is $\propto \Delta$  and double-flip 
$\mu'= -\mu$ which is $\propto \Delta^{2}$ transitions in off-forward CS
with $(\gamma^{*},\gamma^{*'})$ momentum transfer $\vec{\Delta}\neq 0$.
Similar SCHNC would persist also in diffractive 
vector meson production $\gamma^{*}p\to Vp'$ which is obtained from CS
 by continuation from spacelike $\gamma^{*}$ to timelike 
$V$ and swapping the $\Psi^{*}$ for the photon for the vector meson lightcone
wave function \cite{KNZ98,IN99},  for light vector mesons
in the approximation of massless quarks see also \cite{IK}.
The most exciting point about helicity flip in $\gamma^{*}p\to Vp'$ 
is that it uniquely probes spin-orbital angular momentum coupling and Fermi motion 
in vector mesons. 

Experimentally the SCHNC LT-interference in diffractive DIS can be observed
via azimuthal correlation between the $(e,e')$ and 
$(p,p')$ scattering planes. Our work on this asymmetry  and its use for the
determination of $R=\sigma_{L}/\sigma_{T}$ for diffractive DIS has been reported
at DIS'97 \& DIS'98 and published elsewhere \cite{NPZLT}. Here I only
recall that azimuthal asymmetry is the twist-3 effect,
\begin{equation}
A_{LT} \propto {\Delta \over Q}g_{LT}^{D}(\bar{x},\beta,Q^{2})\, ,
\label{eq:2}
\end{equation}
where $g_{LT}^{D}$ is the scaling structure function.  Because  
excitation of $q\bar{q}$ state with $\lambda+\bar{\lambda}=0$ is dual to
production of longitudinal vector mesons, this result for diffractive
 DIS into continuum 
immediately entails that the dominant 
SCHNC effect in vector meson production will be the interference of production 
of longitudinal vector mesons by (SCHC) longitudinal and (SCHNC) transverse photons,
i.e., the element $r_{00}^{5}$ of the vector meson polarization density matrix.

The detailed discussion of the twist and sensitivity of SCHNC amplitudes to
Fermi motion in vector mesons is found in \cite{KNZ98,IN99},  here I only show
in fig.~1 a comparison. between our theoretical estimates \cite{AINP} of $r_{00}^{5}$ for
diffractive production of the $\rho^{0}$ treated as a pure $S$-wave $q\bar{q}$ state 
and the ZEUS \cite{ZEUS} and H1 \cite{H1} experimental data. The agreement
is very good, but much more theoretical work on sensitivity of spin-flip to
short-distance properties of the vector meson wave function is called upon.
Although one would wish to be in the pQCD domain, I emphasize that the existence 
of helicity flip does not require pQCD. In "normal" cases, i.e., helicity non-flip 
and single-helicity flip, the large virtuality of the photon and the form of the 
photon WF ensure the dominance by hard gluon exchange, but the double-flip amplitude
remains dominated by soft gluon exchange, for more discussion see \cite{KNZ98,IN99}.

\begin{figure}[htb]
\vspace{9pt}
\epsfysize 1.6 in 
\epsfbox{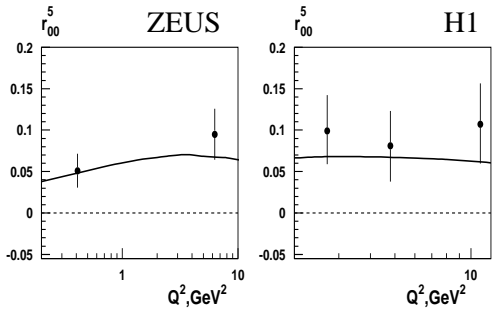}
\caption{Our prediction \cite{AINP} for SCHNC  
spin density matrix element $r_{00}^{5}$ of diffractive $\rho^{0}$-meson vs. the 
data from
ZEUS \cite{ZEUS} and H1 \cite{H1}.}
\label{fig:1}
\end{figure}

In contrast to the $S$-wave state, in the $D$-wave case the total spin 
of $q\bar{q}$ pair is predominantly opposite to the spin of the $D$-wave 
vector meson. This results in a strikingly different structure of helicity-flip 
amplitudes for the $D$ and $S$-wave states, which may facilitate the $D$-wave 
vs. $2S$-wave assignment of the $\rho'(1480)$ 
and $\rho'(1700)$ and of the $\omega'(1420)$ and $\omega'(1600)$,
which remains one of hot issues in the spectroscopy of vector mesons.

Here I only cite the ratio $\rho$ of helicity amplitudes for $S$- and $D$-wave
states normalized to
$V\to e^+e^-$ decay amplitudes \cite{IN99}:
\begin{eqnarray}
\rho_{0L}(D/S)= 
{1\over 5}\left(1-{8m_{V}^{2}\over Q^{2}+m_{V}^{2}}\right)\, ,
\nonumber\\
\rho_{\pm\pm}(D/S)= 
3\left(1+{4\over 15}{m_{V}^{2}\over Q^{2}+m_{V}^{2}}\right)\, ,
\nonumber\\
\rho_{0\pm}(D/S)=-{1\over 5}
(m_{V}a_S)^{2}\left(1+{3m_{V}^{2}\over Q^{2}+m_{V}^{2}}\right)\, ,
\nonumber\\
\rho_{\pm L}(D/S)= 
{3\over 40}(m_{V}a_S)^{4}\, .
\end{eqnarray}
Here $a_{S}$ is the radius of the $S$-wave state and for the 
illustration purposes we used the
harmonic oscillator wave functions. Notice that $A_{0L}^{D}$ changes 
the sign at  $Q^2 \sim 7 m_{V}^{2}$ and the ratio 
$R^{D}=\sigma_{L}/\sigma_{T}$ has thus a non-monotonous $Q^2$ behavior 
and $R^{D} \ll R^{S}$. Second, because $m_{V}a_{S} \gg 1$, for $D$-wave 
states the SCHNC effects are much stronger and the leading SCHNC amplitude
changes the sign from the $S$ to $D$-wave sate. Third, notice the abnormally
large higher twist corrections to the $0L$ and $0\pm$ helicity amplitudes
for $D$-wave state.

The transverse spin asymmetry in polarized DIS is proportional to the
amplitude of forward Compton Compton scattering  
$\gamma^{*}_{L}p\!\!\uparrow \to \gamma^{*}_{T}p\!\!\downarrow$. 
 This Compton 
amplitude and the transverse spin asymmetry 
are proportional to $g_{LT}=g_{1}+g_{2}$. Because 
the photon helicity flip is compensated by the 
target proton helicity flip, the familiar forward zero of this helicity 
amplitude is lifted, but the price one pays for that is that in the 
standard $q\bar{q}$ and/or
two-gluon $t$-channel tower approximation the pomeron exchange does not
contribute to $g_{LT}$ and the transverse asymmetry $A_{2}$ is believed
to vanish at $x\to 0$. The vanishing $A_{2}$ follows also from the
Wandzura-Wilczek relation between $g_{LT}$ and $g_{1}$ \cite{WW}.  

The opening of diffractive DIS channels affects, via unitarity, the
Compton scattering amplitudes. 
In \cite{G2} we have shown how diffractive LT interference in
conjunction with  spin-flip pomeron-nucleon coupling $r_{5}$ give
rise to a manifestly scaling unitarity correction to $g_{LT}$ which rises steeply 
at small $x$,
\begin{equation}
g_{LT}(x,Q^{2}) \propto {1\over x^{2}} r_{5}\int_{x}^{1} {d\beta\over \beta}
g_{LT}^{D}(x_{\Pom}={x\over \beta},Q^{2})\, ,
\end{equation}
and gives rise to the 
transverse spin asymmetry $A_{2} \propto x^{2}g_{LT}$ which does not vanish 
at small $x$. Furthermore, at a moderately small $x$ it even rises because
$g_{LT}^D(x_{\Pom},Q^{2}) \propto G^{2}(x_{\Pom},\overline{Q}^{2})$ where
$\overline{Q}^{2} \sim$ 0.5-1 GeV$^2$. In fig.~2 we show how the unitarity 
correction overtakes at small $x$ the $g_{LT}$ evaluated from the Wandzura-Wilczek 
(WW) relation starting with fits to the world data on $g_{1}$. As such the unitarity
effect is a first nontrivial scaling departure from the WW relation

\begin{figure}[htb]
\vspace{9pt}
\epsfysize 1.8 in 
\epsfbox{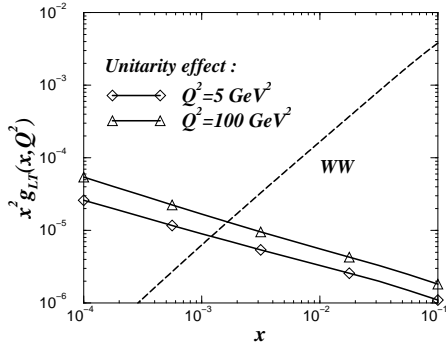}
\caption{The unitarity correction to, and WW relation based evaluation of, $g_{LT}$
\cite{G2}.}
\label{fig:2}
\end{figure}

Notice that neither pomeron nor multipomeron exchanges do contribute to $g_{1}$
and for small $x$ where $g_{LT}$ is dominated by the unitarity correction,
$g_{1} \ll g_{LT}$ and $g_{2}\approx g_{LT}$. Consequently, $g_{2}$ rises at small
$x$ faster that $\propto {1\over x^{2}}$, which invalidates superconvergence 
assumptions behind the derivation of the Burkhardt-Cottingham (BC) sum rule
\cite{BC}. The 
BC integral simply diverges and the BC sum rule does not exist \cite{G2}.

My final comment is on the longitudinal spin asymmetry $A_{L}^{V}$ 
for vector meson production
in polarized DIS on longitudinally polarized protons. The HERMES 
collaboration reported first evidence for this asymmetry in the
$\rho^{0}$ production \cite{HERMES}. 
By the same token of analytic continuation from diagonal Compton scattering  
$\gamma^{*}p\to \gamma^{*}p'$ to $\gamma^{*}p\to Vp'$, this spin asymmetry is
a close counterpart of $A_{1}$ in inclusive polarized DIS. Furthermore, one can
argue that for the starting approximation 
\begin{equation}
A_{L}^{V}\approx 2A_{1}^{q}(\bar{x}),
\end{equation}
the trivial factor 2 being due to the fact that in vector meson production 
one measures the differential cross section, i.e., the amplitude squared,
whereas in inclusive DIS one measures the total Compton cross section 
asymmetries which 
by the optical theorem are linear in the forward Compton scattering amplitude.

Recall that $A_{1}$ receives contributions from quarks
and gluons. Recently it has been argued that contribution from polarized gluons
to $A_{L}^{V}$ is negligible \cite{Mankiewicz}. Consequently, a difference
$A_{1}-{1\over 2}A_{L}^{V}$ is a direct measure of the still elusive 
gluon contribution to $g_{1}$ of the proton!

\end{document}